\documentclass[10pt,conference]{IEEEtran}

\usepackage{hyperref}       
\usepackage[natbib=true, sorting=none]{biblatex}
\usepackage[utf8]{inputenc}
\usepackage[T1]{fontenc}    
\usepackage{amsmath,amsthm,amssymb,amsfonts}       

\usepackage{booktabs} 
\usepackage{graphicx}
\usepackage{physics}
\usepackage[capitalise]{cleveref}
\usepackage{adjustbox}
\usepackage{tikz}
\usepackage{fancyhdr}
\usetikzlibrary{arrows,shapes,positioning,shadows,trees,calc}

\tikzset{
  basic/.style  = {draw, text width=2cm, drop shadow, rectangle},
  root/.style   = {basic, rounded corners=2pt, thin, align=center, fill=blue!10, text width=11em},
  level 2/.style = {basic, rounded corners=2pt, thin, align=center, fill=gray!10, text width=8em, text height=1em},
  level 3/.style = {basic, rounded corners=2pt, thin, align=center, fill=white, text width=11em},
  level 4/.style = {basic, thin, align=left, fill=white, text width=8em}
}

\begin{document}
\title{Predominant Aspects on Security for Quantum Machine Learning: Literature Review}

\fancypagestyle{specialfooter}{%
  \fancyhf{}
  \renewcommand\headrulewidth{0pt}
  \fancyfoot[R]{ \noindent\fbox{%
    \parbox{\textwidth}{%
        {\footnotesize \copyright 2024 IEEE. Personal use of this material is permitted. Permission from IEEE must be obtained for all other uses, in any current or future media, including reprinting/republishing this material for advertising or promotional purposes, creating new collective works, for resale or redistribution to servers or lists, or reuse of any copyrighted component of this work in other works.}
        }
    }}
}

\author{
Nicola Franco\IEEEauthorrefmark{1}, 
Alona Sakhnenko\IEEEauthorrefmark{1}, 
Leon Stolpmann\IEEEauthorrefmark{2},
Daniel Thuerck\IEEEauthorrefmark{3},\\
Fabian Petsch\IEEEauthorrefmark{4},
Annika Rüll\IEEEauthorrefmark{4},
Jeanette Miriam Lorenz\IEEEauthorrefmark{1} \\
\IEEEauthorblockA{\IEEEauthorrefmark{1}Fraunhofer Institute for Cognitive Systems (IKS),  Munich, Germany}
\IEEEauthorblockA{\IEEEauthorrefmark{2}adesso Switzerland, Zurich, Switzerland}
\IEEEauthorblockA{\IEEEauthorrefmark{3}Quantagonia, Bad Homburg, Germany}
\IEEEauthorblockA{\IEEEauthorrefmark{4}Federal Office for Information Security (BSI), Bonn, Germany}
}

\maketitle
\thispagestyle{plain}
\pagestyle{plain}
\thispagestyle{specialfooter}

\begin{abstract}
Quantum Machine Learning (QML) has emerged as a promising intersection of quantum computing and classical machine learning, anticipated to drive breakthroughs in computational tasks. This paper discusses the question which security concerns and strengths are connected to QML by means of a systematic literature review. We categorize and review the security of QML models, their vulnerabilities inherent to quantum architectures, and the mitigation strategies proposed. The survey reveals that while QML possesses unique strengths, it also introduces novel attack vectors not seen in classical systems. We point out specific risks, such as cross-talk in superconducting systems and forced repeated shuttle operations in ion-trap systems, which threaten QML's reliability. However, approaches like adversarial training, quantum noise exploitation, and quantum differential privacy have shown potential in enhancing QML robustness. Our review discuss the need for continued and rigorous research to ensure the secure deployment of QML in real-world applications. This work serves as a foundational reference for researchers and practitioners aiming to navigate the security aspects of QML.
\end{abstract}


\section{Introduction}

In recent years, there has been a marked escalation in quantum technology capabilities, both from the perspectives of engineering and algorithmic design. 
Many global players are joining in on this venture and are developing their own Quantum Computing (QC) stacks. 
Large-scale noise-resilient quantum computers have a proven potential to accelerate the computation of algorithms in many domains~\cite{shor1994algorithms, grover1996fast}, however, these types of computers will not be available in the near future. 
Consequently, much of the current research is centered on harnessing the capabilities of noisy intermediate-scale quantum (NISQ) devices~\cite{preskill2018quantum}.
Machine Learning (ML) is widely believed to have potential to be one of the first applications to benefit from NISQ devices~\cite{lloyd2014quantum, schuld2015introduction, biamonte2017quantum}.
Pioneering efforts in Quantum ML (QML) have led to the development of methods such as quantum clustering~\cite{lloyd2014quantum}, quantum deep learning~\cite{wiebe2014quantum}, and quantum reinforcement learning~\cite{dong2008quantum}.
However, the intersection of QC and ML also spawns unique security challenges, extending beyond conventional ML concerns~\cite{10.1145/3526241.3530833}. 

\medskip

This literature survey examines the security aspects of QML, focusing on both vulnerabilities and defenses. 
The review systematically studies vulnerabilities specific to QML, such as security issues with quantum classifiers in higher dimensions and attack strategies targeting quantum data encodings or leveraging quantum noise.
These findings are important due to the complexity of QML and its potential for practical use, highlighting the need for advanced verification methods to ensure security while achieving quantum advantage.
Additionally, the survey discusses proactive defense methods. 
These methods include adversarial training, adding privacy measures for data protection, verifying model robustness, and considering hardware noise not only as a weakness but also as a benefit for improving QML model robustness.

By providing a comprehensive analysis of both vulnerabilities and defenses, this survey equips researchers and practitioners with a deeper understanding of the challenges and opportunities in securing QML.

\medskip
The paper is organized as follows: Section \ref{sec:introduction} provides an overview of the basic principles of QC and QML.
Section \ref{sec:research_method} presents the approach used for this systematic survey.
Section \ref{sec:vulnerabilities} delves into the unique challenges of QML models.
Section \ref{sec:quantum_defenses} outlines potential defense mechanisms and  the inherent resilience of these models. 
Finally, Section \ref{sec:conclusion} offers guidance for practitioners by highlighting potential security gaps in QML that warrant future exploration.
\section{What is QML?}\label{sec:introduction}
This section provides the necessary background in QC and QML for a reader new to the topic, to understand the context of this paper.

\subsection{Introduction to QC}
\subsubsection{Brief history of computation}
At the heart of modern-day computer science lies the \textit{Turing machine}, an abstract mathematical concept developed by Alan Turing in the 1930s. This idealized version of a programmable computer provides the theoretical basis that allows us to argue about whether a problem can effectively be solved by an algorithm, in other words, its \textit{computability}. 
Turing presented the \textit{universal Turing machine} that is not only capable to simulate any other Turing machine, but captures whether or not a problem is solvable through algorithmic means in general, agnostic to the hardware on which the computation is performed. 
In other words, if there is a problem that has an algorithmic solution there exists a universal Turing machine that can simulate this algorithm (\textit{Church-Turing thesis}). 
Modern classical computers are based on Von Neumann architecture, which is a practical implementation of a theoretical Turing machine and hence are universal. 
If classical computers can simulate anything that is computable, what do quantum computers have to offer? 
The answer to this question might lie with the \textit{efficiency} of a algorithm~\cite{nielsen2000quantum}. 
In the 90's the first big breakthroughs for QC took place with Shor's~\cite{shor1994algorithms} (finding prime factors for an integer) and Grover's~\cite{grover1996fast} (search through unsorted database) algorithms. 
These algorithms provide an exponential or polynomial speed up to the best known classical algorithms, a result that holds true even today.\footnote{Even though this is a common knowledge in the field, technically speaking, the best known classical algorithm Number Field Sieve has a sub-exponential runtime \cite{Crandall2001}, which implies a superpolynominal speed up for Shor's algorithm.}

\medskip
\subsubsection{Principles of QC}
How is computation on a QC different from the computation on the classical machine?
In a QC, the primary unit of information is a \textit{quantum bit (qubit)} $\ket{\psi}$, which can exist in a superposition of two states $\ket{0}$ and $\ket{1}$ simultaneously:
\begin{equation}\label{eq:qubit}
    \ket{\psi} = \psi_1 \ket{0} + \psi_2 \ket{1}.
\end{equation}
This qubit represents a \textit{quantum state} that exists in a complex vector space called \textit{Hilbert space} with dimensionality of $2^n$, where $n$ is the number of qubits.
Unlike classical bits that are strictly 0 or 1, qubits allow quantum computers to inherently parallelize computations. 
However, to obtain a classical output from a quantum computation, qubits are measured (wave function collapse), which outputs $0$ with a \textit{probability} ${|\psi_1|}^2$ and $1$ with a \textit{probability} ${|\psi_2|}^2$. 
In contrast to a classical computation, where binary operations are performed in a sequence to arrive to a solution of a complex tasks, quantum computation happens before wave function collapses during which we can employ different quantum phenomena (etc. entanglement, interference) to manipulate the quantum state of the qubit to e.g. maximize the probability of the solution. 
These manipulations are performed by applying a series of \textit{quantum gates}, which are described by \textit{unitary matrices} $U$ (their complex conjugate $U^{\dagger}$ is their inverse $U^{\dagger}U=I$). 
The unitarity of these operations is important as it preserves the inner product between vectors \cite{nielsen2000quantum}.
With quantum gates we can perform a variety of operations, e.g. performing to a simple NOT operation with a quantum X gate (e.g. $\ket{0} \to \ket{1}$), creating superposition state with quantum Hadamard gate, entangling two qubits together with a quantum controlled NOT (CNOT) gate.

\medskip
\subsubsection{State-of-the-art}
Grover's and Shor's algorithms impose certain requirements on the hardware, such as fault-tolerance, which are not yet possible. Therefore, many research efforts today are dedicated to looking for applications that can leverage QCs of NISQ era. One the big recent breakthroughs was done in 2019~\cite{Arute_2019}, which claimed quantum supremacy on a 63-qubits Google Sycamore superconducting chip. As of today, we have superconducting machines with 433 physical qubits with 1k-qubit machines coming in the near future~\cite{ibm_roadmap}. 
These machine have different hardware architectures, ones of the most popular on the market being superconducting chips (due to their similarity to classical architectural designs) and ion traps (which unlike superconducting QCs can operate at a room temperature).

\subsection{Introduction to QML}
ML promises to be one of the earliest application fields to benefit from early adaptation of QC. Many approaches have already been proposed varying from operating on hard datasets consisting of quantum data, such as molecular states \cite{Dallaire_Demers_2018} and phases of matter \cite{lu2020quantum}, to quantum-inspired fully-classical approaches \cite{lu2021tensor}. 
In this work, we concentrate on a particular subfield of QML that is sometimes referred to as \textit{quantum-assisted ML}. 
This subfield develops technique to \textit{hybridize} classical ML with quantum algorithms. 
In this case, quantum algorithms are often represented as Parametrized Quantum Circuits (PQC)~\cite{Benedetti_2019} that are parameterized by classical parameters, which can be tuned by classical methods during the training process. PCQs can be implemented as a stand-alone quantum model, or be a part for a hybrid quantum-classical architecture which is a preferred approach in the field for implementing more complex and bigger QML models.

\medskip
\subsubsection{Data encoding feature map}

A quantum encoding transforms classical data into quantum states within a Hilbert space.
This transformation is akin to a feature map and plays a significant role in determining their computational strength \cite{Schuld_2019, schuld2018supervised}. 
To incorporate classical data into a quantum system, there are several encoding strategies available, a few of which are briefly outlined here. 

One of the most basic embeddings $\Phi(x)$ is \textit{amplitude embedding}. 
With this embedding we encode a $2^N$-dimensional input into $N$-dimensional (logarithmically compact) quantum state as follows:
\begin{equation}
    \Phi(\mathbf{x}): \mathbf{x} \in \mathbb{R}^{2^N} \to \ket{\psi(\mathbf{x})} = \sum_i^{2^N} \psi_i(x_i)\ket{i},  
\end{equation}
where $\sum_i |\psi_i(x_i)|^2 = 1$. This feature map resembles a linear kernel~\cite{Schuld_2019}. 
Another popular embedding for QML models due to its hardware-efficiency is \textit{angle embedding}. This embedding resembles a cosine kernel and encodes $N$-dimensional input into $N$-dimensional quantum states as follows:
\begin{equation}
    \Phi(\mathbf{x}): \mathbf{x} \in \mathbb{R}^N \to \ket{\psi(\mathbf{x})} = \prod_{k=1}^n( sin(x_k) \ket{0} + \cos(x_k)\ket{1}).
\end{equation}

\medskip
\subsubsection{Model zoo}
Binary classifier algorithms are a stepping-stone for a plethora of different models. A quantum binary classifier consists of a PQC that can either accept quantum data as its input e.g. \cite{liu2020vulnerability, lu2020quantum} or encode classical data into a quantum state e.g. \cite{huang2023certified, liao2021robust, Ren_2022}. The parameters of the PQC can be classically tuned to elicit a desired behaviour of the model. The output of the quantum classifier can be a single measure of an output qubit, which will output a correct class with a high enough probability, alternatively a more common approach is to perform multiple measurements to compute an expectation value of a qubit, which can then be binarized, e.g. in \cite{Ren_2022}.

An immense success of NNs in classical ML, provided the motivation for a similar line of research within QML community. PQC's organised in a layer-wise fashion (inspired by classical feedforward architecture) have been coined \textit{quantum NNs (QNNs)}. The QNN structure has many realizations depending on type of data and context similar to quantum classifiers described above. QNNs have been shown to have higher capacity than their classical counterparts in a certain experimental setup \cite{Abbas_2021}. Other approaches that were inspired by more complex classical architectures, such as quantum convolutional (quanvolutional) neural networks, have also shown promising results \cite{sooksatra2021evaluating}. 

Many other ML pillars have been successfully ported to QML realm, e.g. quantum reinforcement learning \cite{dong2008quantum, skolik2023robustness}, quantum generative adversarial networks \cite{Dallaire_Demers_2018, rudolph2022generation}, quantum Boltzmann machines \cite{Zoufal_2021, kehoe2021defence}, quantum kernel methods \cite{Huang_2021}.

\medskip
\subsubsection{Hardware noise}

In quantum computing, \textit{noise} denotes unwanted disturbances in quantum states that might result in computational errors, arising from environmental interactions, quantum gate inaccuracies, or fabrication flaws~\cite{nielsen2000quantum, preskill2018quantum}.
Viewing noise as an intentional adversarial act in quantum devices means deliberately harnessing or exploiting these disturbances to disrupt quantum computation results, much like fault-injections in classical computing mislead ML models~\cite{liu2020vulnerability}.

Examples of adversarial quantum noise include \textit{bit-flip} and \textit{phase-flip} errors, which alter qubit states or their relative phases~\cite{devitt2013quantum}; \textit{depolarization}, making all qubit outcomes equally probable~\cite{lidar2013quantum}; \textit{decoherence}, destroying quantum interference~\cite{zurek2003decoherence}; \textit{control errors} from quantum gate inaccuracies~\cite{chow2009randomized}; \textit{measurement errors} that misread qubits~\cite{riste2015detecting}; \textit{cross-talk} errors, where qubit operations unintentionally influence another, including between different circuits on the same hardware~\cite{gambetta2017building}; manipulation of \textit{shuttle operations} for Ion trap devices, where ions are physically shuttled from one ion trap to another~\cite{kielpinski2002architecture}; and physical disruptions like temperature changes or stray fields~\cite{krantz2019quantum}. 
Recognizing noise as a potential adversarial manipulation deepens the challenge of ensuring quantum system security. 
Consequently, quantum information scientists are rigorously investigating methods to identify and counter such adversarial noise, aiming to bolster the reliability and security of quantum computations~\cite{kundu2022security}.

\section{Research Methodology}\label{sec:research_method}

\begin{figure*}[!ht]
 \centering  
 \includegraphics[trim={0 3.25cm 0 3.25cm},clip,width=0.85\textwidth]{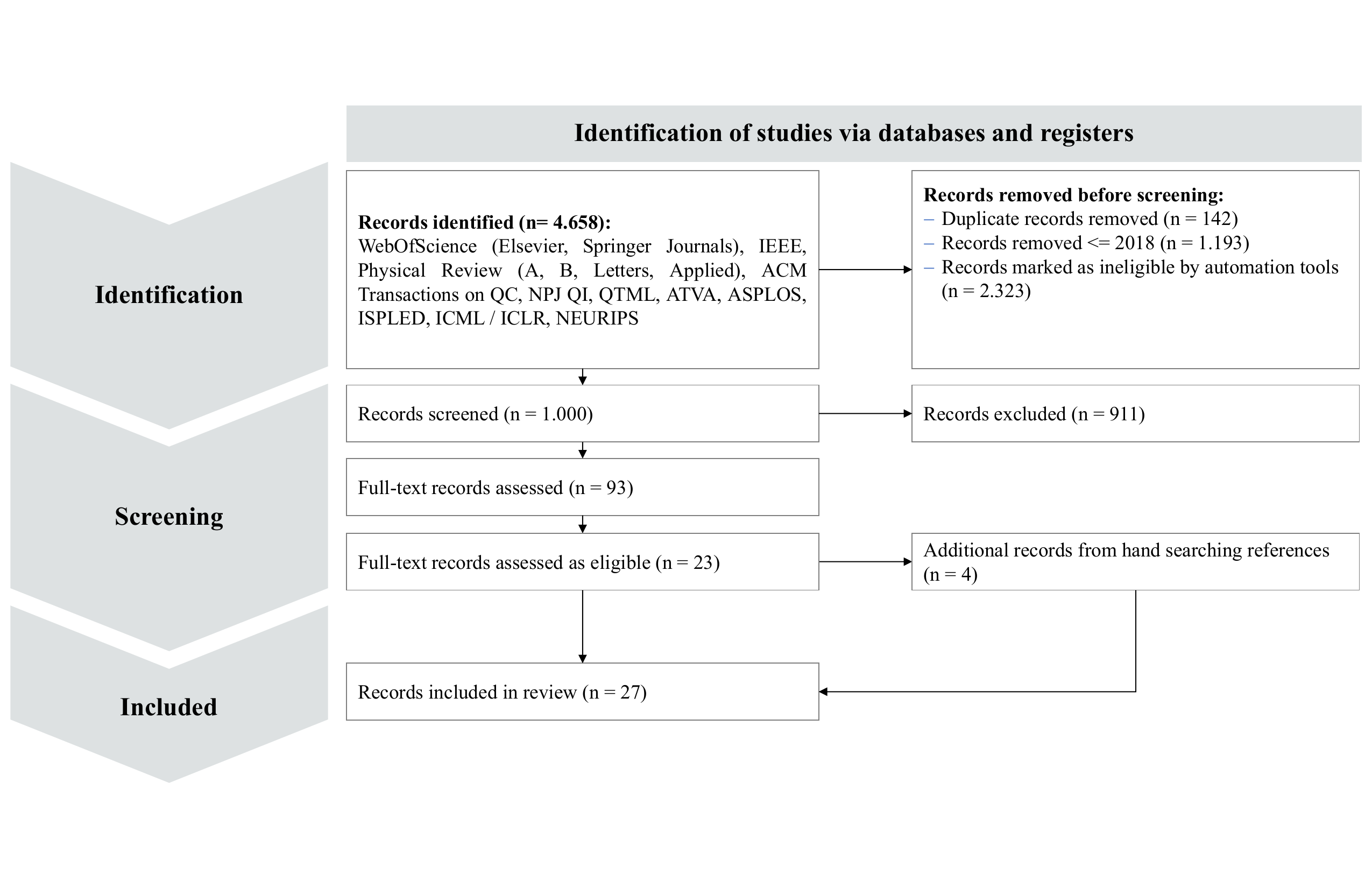}
 \caption{Documentation of search process, adapted from \citeauthor{prisma}~\cite{prisma}}
 \label{fig:PRISMA}
\end{figure*}

In order to prevent information bias, a broad and systematic driven search of academic bibliographic data sources was conducted using the PRISMA~\cite{prisma} framework as an underlying guideline. 
The literature review process was performed by utilizing two search methods executed in the following order and depicted in~\autoref{fig:PRISMA}: 
\begin{itemize} 
    \item[(1)] Execution of the systematic part using the academic bibliographic data sources as depicted in Table \ref{tab:syslitsources} with the search terms as depicted in Table \ref{tab:syslitterms}.
    \item[(2)]  A narrative part that was driven by snowballing through the reference list of selected literature in part (1). 
\end{itemize}
While the bibliography files for most sources could be obtained via API or web site scraping automatically, the search on the arXiv had to be performed manually in a heuristic manner due to the lack of an API and the sheer volume of items, severely restricting the scraping ability.

\begin{table}[h!] 
    \centering
    \caption{Overview of sources for literature research}
    \label{tab:syslitsources}
    \vspace{-1em}
    \begin{adjustbox}{width=0.5\textwidth, center}
     \begin{tabular}{l l l} 
        \toprule
       Databases & Journals & Conferences \\
       \midrule
       Arxiv          & IEEE                      & QTML \\ 
    WebOfScience   & Springer Journals         & IEEE Quantum Week \\
    Elsevier       & Physical Review A/B       & ATVA \\
                   & Physical Review Letters   & ASPLOS\\
                   & Physical Applied          & ISPLED \\ 
                   & ACM Transactions on Quantum Computing & ICML/ICLR \\ 
                   & Nature                    & NeurIPS    \\
     \bottomrule
    \end{tabular}
    \end{adjustbox}
\end{table}

\begin{table}[h!]
    \centering
    \caption{Overview of search terms leading to inclusion into the candidate list}
    \label{tab:syslitterms}
    \vspace{-1em}
    \begin{adjustbox}{width=0.5\textwidth, center}
    \begin{tabular}{l l l l l l} 
       \toprule
        Secur*         & Quantum   & Vulnerabil*   &  Robust*    \\  
        Threat         & Inject*   & Poison*       &  Backdoor   \\ 
        Privacy        & Defens*   & Confidential* &  Encrypt*   \\
        Model stealing & Model reconstruction      &  Crosstalk  & Fault injection     \\
        Encoding       & Adversarial attack        &  Verificat* & Adversarial training\\
        Readout sensing& Attack    & Malicious     &  Certificat*\\ 
        Shadow training \\
       \bottomrule
    \end{tabular}
    \end{adjustbox}
\end{table}

\begin{table}[h!]
    \centering
    \caption{Overview of search terms leading to exclusion from the candidate list}
    \label{tab:excluded}
    \vspace{-1em}
    \begin{adjustbox}{width=0.5\textwidth, center}
        \begin{tabular}{l l l} 
           \toprule 
        Cryptography     & Key Distribution   & Key Exchange    \\
        Encryption 
        & Random Number Generator     & Quantum Communication   \\
        Signcryption   &   Signature  & Post-Quantum     \\
        Blockchain   & Quantum key   &   Cryptanalysis    \\
        \bottomrule  
        \end{tabular}
    \end{adjustbox}
\end{table}

\begin{table}[h!]
    \centering
    \caption{Most important terminology for algorithmic scoring}
    \label{tab:topterms}
    \vspace{-1em}
    \begin{tabular}{l l l l} 
       \toprule  
    Quantum     & Machine Learning   & Secur*   &   Attack \\  [1ex] 
       \bottomrule
    \end{tabular}
\end{table}

For the selection of literature, we included journal and conference papers that specifically addressed the use of quantum computing in the area of ML, with a particular focus on the security aspects and potential attack vectors of these applications. 
Additionally, these papers were required to be written in English. 
On the other hand, we excluded several types of papers: those that merely used our search terms but did not directly relate to QML (detailed in \autoref{tab:excluded}), those not available in electronic format, inaccessible papers, and papers that primarily focused on topics such as \textit{communication}, \textit{key distribution}, and \textit{block-chain}.

\subsection{Filtering \& Scoring}

Upon execution of the systematic search as depicted above $4\,658$ records were identified. 
In these, $142$ duplicates were identified and removed. 
Further, records before year 2018 were removed ($1\,193$) and records marked as ineligible by the automation tool amounting to $2\,323$ records were removed. 
After their removal, $1\,000$ records were left before the screening phase. 

The automation tool used the title and abstract for each record inputs to compute a score of relevance. First, it assigns a score to each query of \autoref{tab:syslitterms}. 
Since the list is roughly sorted from general to specific, the score equals first twice the position of the term in the list as a hit (e.g. Secur* -> 2, Quantum 4, ...). These points are each doubled when the term is paired with \textit{Quantum}. Finally, we add another 200 points if one of the combinations of terms depicted in~\autoref{tab:topterms} was identified or deduct 300 points if any of the search terms depicted in~\autoref{tab:excluded} appear. 
The penalties were set iteratively through repeated manual analysis of the Top 50 results given a set of penalties.

In total, $1\,000$ records remained for the screening phase. Of these we excluded $911$ because of exclusion criteria listed above. The remaining $92$ records were screened taking title and abstract for eligibility leaving $25$ records that were included in the review with addition $7$ records retrieved from references of the full-text screened records. Additionally, we excluded the literature that required the physical manipulation of the hardware or attack strategies that had a direct classical analogue $9$, but added additional $4$ more in-depth theoretical papers from discarded works that resurfaced during the writing of this work. 
In total $27$ records are included in the literature research results.

\section{Quantum Vulnerabilities}\label{sec:vulnerabilities}

\begin{figure*}[h]
    \centering
    \begin{minipage}{\textwidth}
    \centering
    \begin{tikzpicture}[
      level 1/.style={sibling distance=40mm},
      edge from parent/.style={->,draw},
      scale=1., every node/.style={scale=1., font=\footnotesize},
      >=latex]
    \node [root] {QML Vulnerabilities}
    child {node [level 2, xshift=-30] (attack) {Attack Vectors}
      child [grow=down] {node [level 3, below of = attack, xshift=-30mm, node distance=1cm] (fault) {Fault Injections~\cite{chu2023qtrojan, chu2023qdoor}} edge from parent[draw=none]} 
      child [grow=down] {node [level 3, below of = attack, xshift=15mm, node distance=1cm] (noise) {Quantum Noise~\cite{saki2021impact, saki2021shuttleexploiting, 9794194, 8824907}
      } edge from parent[draw=none]}
    }
    child {node [level 2, xshift=30] (scaling) {Scaling Pitfall} 
        child [grow=down] {node [level 3,  below of = scaling, node distance=1cm] (sensitivity) {Sensitivity~\cite{liu2020vulnerability, gong2022universal}} edge from parent[draw=none]}
    };
    
    \draw[->] (attack) -- (fault);
    \draw[->] (attack) -- (noise);
    \draw[->] (scaling.south) -- (sensitivity.north);

    \end{tikzpicture}
    \end{minipage}%
    \caption{A taxonomy diagram detailing the key areas of vulnerabilities in QML.
    }
    \label{fig:attack_taxonomy}
\end{figure*}
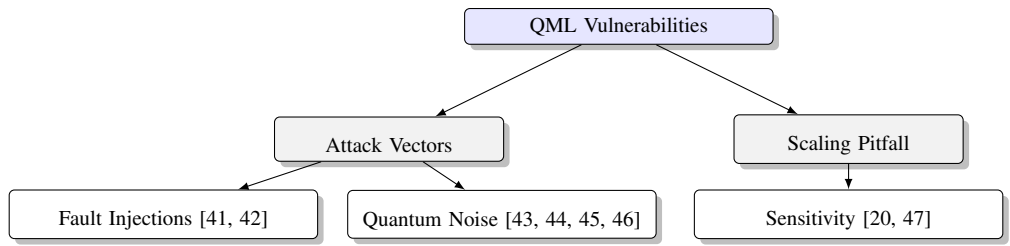

Here, we compile vulnerabilities and attack vectors that are unique for QML models and discuss possible mitigation strategies. This topic is subdivided into two parts as follows:

\begin{enumerate}
    \item \textit{Quantum attack vectors} describe a way to exploit specificities of QML models to craft an attack that has no classical analog.
    \item \textit{Scaling pitfall} characterizes a trait of larger quantum classifiers that might introduce a trade-off between system security and quantum advantage.
\end{enumerate}

These vulnerabilities pose a new challenge for secure utilization of QML methods.
\subsection{Quantum attack vectors}\label{sec:attack_vectors}

In this section, we review attack vectors of QML models, encompassing those without direct classical analogues. 
One of the most prominent types of this kind of attacks are \textit{fault injections}, which we review in detail below.

\medskip
\subsubsection{Fault Injections}

A \textit{quantum Trojan virus} can function as a backdoor access to a QNN architecture that would allow for targeted gate injection as indicated by \citet{chu2023qtrojan}. This work showed that when the virus is dormant, the model will achieve the same performance as in its unaltered state making compromised model hard to detect. However, when the virus is activated introducing perturbation of an encoding feature map, it leads to nearly 100\% attack success rate. The authors hinted that classical methods against Trojan infiltration might be a viable defense in this case, however, it requires further investigation.
In this direction, \citet{chu2023qdoor} exploit the unitary differences between uncompiled and synthesized QNN circuits, embedding backdoors that activate post-synthesis. 
This strategy drastically improves the attack's success rate while preserving high clean data accuracy, highlighting the method's effectiveness and stealthiness over existing approaches.

\medskip
\subsubsection{Exploiting quantum noise}

In a co-tenancy setting, where multiple users simultaneously execute quantum programs on the same machine, an attacker can craft a program that would induce hardware-dependent errors which could lead to the \textit{degradation of performance} of quantum models or induce \textit{denial-of-service (DoS)} attack.
In case of superconducting systems, an attacker can induce cross-talk between quantum circuits~\cite{saki2021impact}, while, in case of ion-trap systems, an attacker can induce repeated shuttle operations~\cite{saki2021shuttleexploiting} on shared qubits - both of which decrease the reliability of quantum circuits. 
Moreover, perpetrator can exploit correlating read-out errors to perform an \textit{information leakage} attack~\cite{saki2021impact}.

We can tackle these issues head on by isolating user programs by introducing buffering / padding qubits~\cite{9794194, saki2021shuttleexploiting} or by introducing circuit-pattern recognition methods into quantum clouds to spot error inducing circuits~\cite{9840181}. 
Another approach is to borrow the methodologies designed for increasing the reliability of QML during inference. 
Quantum hardware vendors periodically re-calibrate their devices, which means executing identical PQCs on the same hardware at different times might yield erratic results~\cite{8824907}. 
A few mitigation techniques have been proposed in the literature, e.g. embodying noise into the training loop of a PQCs~\cite{8824907, saki2021shuttleexploiting} or by exploiting just-in-time compilation technique~\cite{wilson2020justintime, 9794194}.

\subsection{Scaling Pitfall}

We expect to attain quantum advantage with bigger quantum systems that span higher dimensional quantum Hilbert spaces. However, as the dimension grows, sensitivity of a quantum classifier to minor perturbations near the decision boundary increases, making quantum classification vulnerable and demanding more resources for verification as shown by \citet{liu2020vulnerability}. This sensitivity demands an exponential increase in resources to verify security, potentially offsetting quantum advantages. 
Similarly, \citet{gong2022universal} indicated the existence of universal adversarial perturbations in higher dimensions, meaning a minor adversarial perturbation of the input is sufficient to pose a moderate risk to a whole set of quantum classifiers. 
Drawing from the quantum no-free-lunch theorem~\cite{poland2020no}, they noted that this risk increases exponentially with the classifier's qubit count. 
Their findings emphasize the need to address adversarial threats in quantum data classification and call for experimental validations beyond the theoretical study.

\section{Quantum defenses}\label{sec:quantum_defenses}

Here, we summarize available defenses to attacks on QML models. In this context, several defenses are emerging to counter potential threats, which we collect under three main areas: \textit{adversarial training}, \textit{differential privacy} and \textit{formal verification}.

\begin{enumerate}
    \item \textit{Adversarial training} enhances quantum models by exposing them to deliberately crafted malicious inputs, thereby improving resilience.
    \item \textit{Differential privacy} offers a probabilistic approach that introduces noise to the data to achieve more resilient predictions.
    \item \textit{Formal verification} refers to the process of rigorously proving the correctness of quantum algorithms and models using mathematical methods.
\end{enumerate}

Together, these strategies are forging a secure path forward for QML applications.

\begin{figure*}[h]
    \centering
    \begin{minipage}{\textwidth}
    \centering
    \begin{tikzpicture}[
      level 1/.style={sibling distance=40mm},
      edge from parent/.style={->,draw},
      scale=1., every node/.style={scale=1., font=\footnotesize},
      >=latex]
    \node [root] {QML Defenses}
    child {node [level 2, xshift=-30] (training) {Adversarial Training}
      child [grow=down] {node [level 3, below of = training, node distance=1cm] (gradient) {Gradient-based~\cite{lu2020quantum, liu2020vulnerability,  banchi2022robust, gong2022universal, akter2023exploring, West_2023}} edge from parent[draw=none]} 
      child [grow=down] {node [level 3, below of = gradient, node distance=1cm] (architecture) {Resilient Architectures~\cite{sooksatra2021evaluating, yang2023improved, gong2022enhancing}
      } edge from parent[draw=none]}
    }
    child {node [level 2, xshift=0] (privacy) {Data Privacy} 
        child [grow=down] {node [level 3,  below of = privacy, node distance=1cm] (device) {Differential Privacy~\cite{du2021quantum, liao2021robust, Weber_2021, huang2023certified, hirche2023quantum, sahdev2023adversarial}} edge from parent[draw=none]}
        child [grow=down] {node [level 3, below of = device, node distance=1cm] (adversarial) {Inherent Privacy~\cite{liao2021robust, kumar2023expressive}} edge from parent[draw=none]}
    }
    child {node [level 2, xshift=30] (formal) {Formal Verification} 
        child [grow=down] {node [level 3,  below of = formal, node distance=1cm] (randomized) {MILP Verification~\cite{franco2022quantum, franco2023efficient}} edge from parent[draw=none]}
        child [grow=down] {node [level 3, below of = randomized, node distance=1cm] (exact) {Lipschitz~\cite{barooti2021provable, guan2020robustness, guan2021robustness, guan2022verifying}} edge from parent[draw=none]}
    };
    
    \foreach \node in {gradient, architecture} {
        \draw[->] (training.west) -- ($(training.west) + (-1,0)$) |- (\node.west)+(-0.2,0) -- (\node.west);
    }

    \foreach \node in {device, adversarial} {
        \draw[->] (privacy.west) -- ($(privacy.west) + (-1,0)$)  |- (\node.west)+(-0.2,0) -- (\node.west);
    }

    \foreach \node in {randomized, exact} {
        \draw[->] (formal.west) -- ($(formal.west) + (-1,0)$)  |- (\node.west)+(-0.2,0) -- (\node.west);
    }

    \end{tikzpicture}
    \end{minipage}%
    \caption{A taxonomy diagram detailing the key areas and methodologies within the domain of defenses for QML. 
    }
    \label{fig:defense_taxonomy}
\end{figure*}
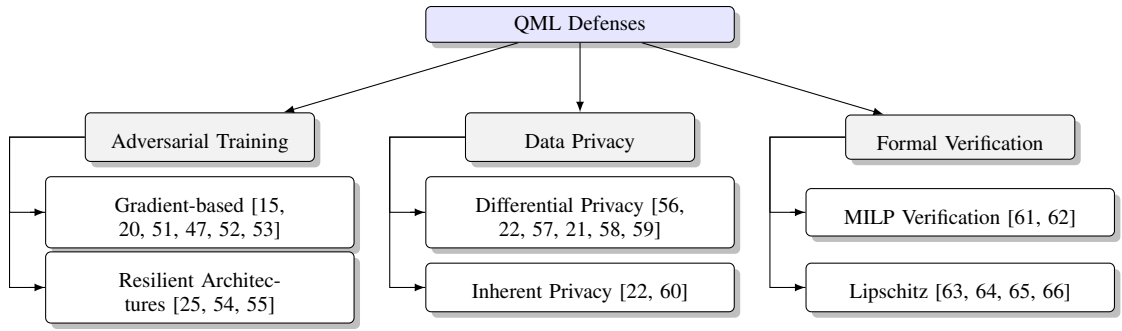

\subsection{Adversarial Training}

One main practice in adversarial ML is to train the model with adversarial attacks, enhancing its robustness against such threats.
Adversarial attacks manipulate inputs to maximize model errors, often in ways imperceptible to humans.
Common classical approaches are: (i) Fast Gradient Sign Method (FGSM)~\cite{fgsm} a straightforward method that perturbs inputs using the sign of the gradient to maximize model error, (ii) Basic Iterative Method (BIM)~\cite{bim} an enhanced FGSM that iteratively applies perturbations, (iii) Projected Gradient Descent (PGD)~\cite{pgd} an iterative method that ensures adversarial examples stay within a valid, imperceptible range, (iv) Momentum Iterative Method (MIM)~\cite{dong2018boosting} which incorporates momentum to enhance the attack by evading local maxima.

\medskip
\subsubsection{Classical techniques adapted to QML}

Early research by \citet{lu2020quantum} highlighted the susceptibilities of QML to established adversarial attacks like PGD~\cite{pgd}, FGSM~\cite{fgsm}, MIM~\cite{dong2018boosting} and BIM~\cite{bim}. 
The study revealed that quantum classifiers on both classical and quantum data can fall victim to adversarial tactics, much like traditional classifiers. 
However, employing adversarial training can bolster their defenses considerably. 
Similarly, \citet{Ren_2022} highlighted that adversarial training enhances the resilience of variational quantum classifiers on a superconducting QC. 
Using classical optimization in a white-box setting, they produced adversarial examples that led to misclassifications. 
When these samples were integrated into the training process, the classifiers successfully resisted adversarial attacks.

\medskip
\subsubsection{Techniques for enhanced quantum adversarial robustness}

Another line of research explores the foundational principles and techniques for enhancing the robustness of quantum systems against adversarial threats.
In particular, \citet{West_2023} investigate the concept of attack transferability, where adversarial samples from one network are used to attack similar networks. 
Interestingly, quantum variational circuits appear more resistant to these transferred attacks compared to classical networks. 
This suggests that quantum classifiers might capture more robust data features than classical methods.
Additionally, while adversarial training can enhance resilience, its benefits are reduced for quantum classifiers that are already somewhat resistant. 
In the end, the study proposes that as quantum classifiers scale up, they may outperform classical ones not just in performance but also in robustness, identifying robustness as a potential key strength of quantum classifiers.
\medskip

Recently, \citet{hou2023quantum} presented a quantum adversarial metric learning approach that efficiently computes the triplet loss function, conserving quantum resources. 
They craft a technique for generating adversarial samples via quantum gradient ascent, mirroring the classical approach seen in \cite{Ren_2022}. 
Using these samples in simulated adversarial training, they found improved resilience against adversarial attacks.

Another approach was introduced by \citet{gong2022enhancing}, in which the authors proposed a  random unitary encodings strategy that showed a notable increase in the adversarial robustness of a quantum classifier. This work offers a rigorous proof and empirical evidence of the resilience of this method against white-box and black-box adversarial attacks.

\medskip
\subsubsection{Resilient architectures}

Studies highlighted the resilience of quantum-enhanced architectures in facing adversarial attacks.
Specifically, \citet{kehoe2021defence} demonstrated the superior robustness of Restricted Boltzmann Machines (RBMs) against adversarial attacks, particularly when compared to conventional neural networks.
Their research further introduced quantum-enhanced RBMs, achieving either comparable or marginally superior results on a scaled-down MNIST dataset~\cite{lecun-mnisthandwrittendigit-2010}.
Similarly, \citet{sooksatra2021evaluating} investigated the benefits of incorporating quantum layers into neural networks. 
Their findings indicate heightened accuracy and improved resilience against adversarial threats, with particular emphasis on the effectiveness of \textit{quanvolutional} layers. 
On a parallel note, \citet{suryotrisongko2022adversarial} developed a deep learning model enriched with quantum computing methodologies, specifically tailored to detect and combat botnet Domain Generation Algorithm (DGA) threats in contemporary 5G environments, emphasizing the role of adversarial training in bolstering system robustness.

\subsection{Data Privacy}
The importance of safeguarding sensitive training datasets has become more imperative in the last years as more and more companies are reporting data breaches. Many ways to ensure privacy of the data have been proposed in both the ML and QML communities. 

\medskip
\subsubsection{Differential Privacy}

Differential privacy is a pioneering privacy framework tailored to protect individual data within datasets. 
Introduced by \citet{dwork2006differential}, this mechanism adds controlled noise to computations, allowing models to generalize without compromising privacy. 
In the context of ML, it ensures that a model, trained on a dataset, does not inadvertently reveal sensitive information about any single data point and maintains stable predictions~\cite{abadi2016deep}.

\medskip

\citet{huang2023certified} conducted the first theoretical study on how quantum rotation noise, analogous to classical randomized smoothing~\cite{cohen2019certified}, boosts the resilience of quantum classifiers against adversarial attacks. 
Using the quantum differential privacy framework from \citet{zhou2017}, they proved that a quantum classifier's privacy level, inversely relates to the noise level. 
Their research revealed that increasing rotation noise during training directly increases the classifier's robustness.
In a similar work, \citet{Weber_2021} delve into the link between Quantum Hypothesis Testing (QHT) and a quantum classifier's robustness against unanticipated noise, establishing a vital robustness criterion. 
Their work provides certification protocols for optimal robustness, methods to validate classification results amidst noise, and robustness conditions for different noise types, all articulated using QHT terminology. 
They also hint at an interesting balance between accuracy and robustness in quantum classification.
The defense works by aggregating a classifier’s output in a region around the input and computing the average probability of a class via Monte Carlo estimations. 
\citet{sahdev2023adversarial} employ quantum phase estimation, achieving a quadratic speedup in achieving adversarial robustness compared to classical ML methods, which typically necessitate between $10^5$ to $10^6$ samples for randomized smoothing.

\medskip

\citet{du2021quantum} showed that adding depolarization noise to quantum circuits used for classification can improve their resilience against adversarial attacks, introducing a concept akin to quantum differential privacy. This enhanced robustness is primarily influenced by the number of classes and not the specific classification model, suggesting a potential edge for quantum over classical models. 
Lastly, \citet{liao2021robust} scrutinize the robustness across random pure states and contrasts it with their prior error-region robustness work. 
They assert that vulnerabilities' real-world implications are constrained since adversarial risk should consider a subset of relevant states. 
They also suggest a technique for gauging the resilience of quantum classifiers with states derived from a Gaussian latent space, finding the robustness declines only marginally with increasing encoded qubits.

\medskip
\subsubsection{Inherent Privacy}

In a new line of research, \citet{kumar2023expressive} investigates properties of overparametrized QML models with highly-expressive (in terms of Fourier frequency spectrum \cite{Schuld_2021}) encoding that provide an inherent protection against gradient inversion attacks. The authors examined these models in the context of a federated learning setup, which is a distributed approach to train NNs without the need to share sensitive data with an aggregator. They showed theoretical results backed by empirical evidence that the perpetrator encountered a considerable challenge when attempting to craft an attack on these models due to its complexity. Overparameterization of the trainable PQC ensures a training process that is free of spurious local minima issues.

\subsection{Formal Verification}

To defend against unidentified adversaries, it is essential to formally ensure robustness in worst-case scenarios, which often differ from typical known noise sources like depolarization noise. 
\citet{barooti2021provable} suggest that quantum adversarial robustness hinges on two key conditions: having a generative model to capture sample distributions and a classifier resilient to data distribution shifts. 
By meeting these conditions, we can guarantee robustness. 
Using a prover/verifier quantum model, they explore protocols ensuring adversarial robustness for various classifiers, including quantum and classical ones. 
The study uncovers links between certifiable randomness, generative models, robustness, and cryptographic protocols, prompting questions about applying these findings in real-world defenses against adversarial attacks.

\medskip
\subsubsection{MILP Verification}

\citet{franco2022quantum} introduce a pioneering method for verifying the robustness of ReLU neural networks using quantum computing, transforming the verification challenge into a mixed-integer linear program (MILP). 
They employ Benders decomposition to split the MILP into a quadratic unconstrained binary optimization (QUBO) and a linear program (LP), with the QUBO addressed through variational quantum algorithms or quantum annealing. 
This marks one of the initial efforts to use quantum computers for neural network certification. 
Recognizing the computational demands of traditional MILP approaches for large networks, \citet{franco2023efficient} further explore the efficiency of Benders and Dantzig-Wolfe decompositions in quantum contexts, finding that the Dantzig-Wolfe decomposition significantly lowers qubit requirements, offering a more efficient pathway for leveraging both classical and quantum computing resources in neural network verification.

\medskip
\subsubsection{Lipschitz Continuity}

Leveraging Semi-definite programming, \citet{guan2020robustness, guan2021robustness} derived an analytical measure for robustness that can be practically calculated, offering a practical lower bound of robust accuracy. 
Additionally, they formulated a verification algorithm to precisely confirm the $\epsilon$-robustness of QML models, presenting valuable counter-examples beneficial for adversarial training.
In a subsequent study, \citet{guan2022verifying} introduce a framework for assessing fairness in QML decision models. By leveraging the trace distance to measure data similarity, they link the fairness verification process to distinguishing quantum measurements. Interestingly, they also uncover that quantum noise can potentially enhance model fairness.
\section{Discussion \& Future Directions}

The exploration of vulnerabilities and defenses in QML presents a fertile ground for innovative research. 
Specifically, the examination of vulnerabilities reveals critical areas for future research and improvement.

\medskip
\paragraph{New Attack Vectors}
Research into fault injections and the manipulation of quantum noise illustrates the complex interplay between quantum system architecture and potential attack vectors. 
Particularly, showing the necessity for a deeper understanding of how QML models can be compromised, emphasizing the need for security measures that go beyond classical analogs. 
The scenario becomes even more complex with the increasing scale, where the expansion of quantum systems introduces heightened sensitivity to perturbations. 
This scaling issue brings to light the delicate balance between achieving quantum advantage and maintaining system security, urging a redirection of focus towards developing robust verification methodologies capable of keeping pace with the expansion of Hilbert spaces.

\medskip
\paragraph{Enhancing QML Security}
On the defense front, the areas of adversarial training, differential privacy, and formal verification offer promising avenues for fortifying QML systems against attacks. 
The adaptation of classical adversarial training techniques to the quantum domain offers a foundational step towards enhancing quantum model resilience. However, the unique aspects of QC demand that these techniques be further refined and tailored to effectively counter quantum-specific threats. Differential privacy emerges as a crucial strategy in safeguarding data privacy and integrity, yet the implementation of such methods in quantum environments remains in its nascent stages, calling for more rigorous exploration and adaptation.
On the one hand, MILP verification and Lipschitz continuity offer a solution for mathematically rigorous defenses against quantum adversarial attacks.
On the other hand, these methods are still in the early stages of application to quantum systems and necessitate significant development to be widely applicable and effective.

\subsection{Cross-Disciplinary Research}
The intersection of classical and QML defenses offers unexploited potential. Research into hybrid defense mechanisms that leverage the strengths of both domains can lead to innovative solutions that are robust against a wide array of attacks. However, there is a critical need for empirical studies and benchmarking frameworks that evaluate the effectiveness of proposed defenses under realistic conditions. 
This includes the development of standardized datasets and evaluation metrics specific to QML vulnerabilities and defenses.
In summary, the future of QML security research lies in a multidisciplinary approach that bridges QC and ML, leveraging insights from each to address the unique challenges presented by QML. 
By focusing on these key areas, the research community can advance towards creating QML models that are not only powerful and efficient but also secure and resilient against adversarial threats.
\section{Conclusion}\label{sec:conclusion}

QML models have unique security challenges that need to be addressed before they can be safely deployed into production environments. 
A malicious actor could compromise the architectural integrity or manipulate and utilize inherent noise in a way that is both damaging and difficult to detect.
Some defense mechanisms have been proposed in literature, but require further empirical studies to determine their reliability and effectiveness. 
On the other hand, quantum models are showing resilience against classical attack vectors. 
This strength can be attributed to multiple factors: (i) successful adaptation of both classical and quantum techniques for adversarial training of QML models; (ii) inherent resilience of certain architecture, such as quanvolutional networks and quantum-enhanced RBMs, towards adversarial attacks; (iii) utilization noise inherent to the devices to boost differential privacy; or (iv) the employment of models that inherently protect sensitive data from gradient inverse attacks.

\medskip
In conclusion, the future of QML security research hinges on a multidisciplinary strategy that integrates QC and ML, leading into unexplored potential through hybrid defenses. This approach requires empirical studies and new benchmarking frameworks, alongside the creation of standardized datasets and evaluation metrics tailored to QML-specific vulnerabilities and defenses, aiming to develop QML models that are efficient, powerful, secure, and resilient against diverse adversarial threats.

\printbibliography

\end{document}